\documentclass[12pt]{article}

\usepackage{amsmath}
\usepackage{amsfonts}
\usepackage{amssymb}
\usepackage{graphicx}
\usepackage{subfigure}
\usepackage{rotating}
\usepackage{color}
\usepackage{amsthm}
\usepackage{authblk}
\usepackage{natbib}
\usepackage{threeparttable}
\usepackage{pdflscape}
\usepackage{multirow}
\usepackage{accents}
\usepackage{longtable}
\usepackage{setspace}
\usepackage{url}
\usepackage{mathrsfs}
\usepackage[nolists]{endfloat}

\allowdisplaybreaks[3]

\def \bfr {\mathbf{r}}

\def \tr  {\text{tr}}

\begin{document}
\setlength{\textheight}{575pt}
\setlength{\baselineskip}{23pt}

\title{A Penalized Multi-trait Mixed Model for Association Mapping in Pedigree-based GWAS}
\author[1]{Jin Liu$^{\dagger}$}
\affil[1]{School of Public Health, Yale University, New Haven, CT 06520, U.S.A.}
\author[1]{Can Yang$^{\dagger}$}
\author[1,2]{Xingjie Shi}
\affil[2]{School of Statistics and Management, Shanghai University of Finance and Economics, Shanghai, China}
\author[1]{Cong Li}
\author[3]{Jian Huang}
\author[1]{Hongyu Zhao$^*$}
\affil[3]{Department of Statistics $\&$ Actuarial Science, Department of Biostatistics, University of Iowa, Iowa City, IA 52242, U.S.A.}
\author[1]{Shuangge Ma$^*$}

\maketitle
\def\ep{\varepsilon}
\def\ba{{\boldsymbol a}}
\def\bA{{\boldsymbol A}}
\def\cA{{\cal A}}
\def\hA{\widehat{A}}
\def\tba{\widetilde{\boldsymbol a}}
\def\bb{{\boldsymbol b}}
\def\hb{\hat{b}}
\def\hbb{\hat{\boldsymbol b}}
\def\tbb{\tilde{\boldsymbol b}}
\def\barb{\bar{b}}
\def\bbb{\overline{\bb}}
\def\Bbar{\overline{B}}
\def\bB{{\boldsymbol B}}
\def\cB{{\cal B}}
\def\hcB{{\widehat{\cal B}}}
\def\bC{{\boldsymbol C}}
\def\bD{{\boldsymbol D}}
\def\bd{{\boldsymbol d}}
\def\be{{\boldsymbol e}}
\def\cE{{\cal E}}
\def\rE{{\mathrm E}}
\def\bff{{\boldsymbol f}}
\def\hf{\widehat{f}}
\def\cF{{\cal F}}
\def\bg{{\boldsymbol g}}
\def\bG{{\boldsymbol G}}
\def\cG{{\cal G}}
\def\bh{{\boldsymbol h}}
\def\cH{{\cal H}}
\def\bI{{\boldsymbol I}}
\def\bell{{\boldsymbol \ell}}
\def\tM{\widetilde{M}}
\def\bO{{\boldsymbol O}}
\def\cO{{\cal O}}
\def\bp{{\boldsymbol p}}
\def\rP{{\mathrm P}}
\def\bP{{\boldsymbol P}}
\def\bbP{{\mathbb P}}
\def\tP{\widetilde{P}}
\def\bQ{{\boldsymbol Q}}
\def\bfr{{\boldsymbol r}}
\def\tr{\tilde{r}}
\def\hr{\hat{r}}
\def\hbr{\hat{\boldsymbol r}}
\def\tbr{\tilde{\boldsymbol r}}
\def\bs{{\boldsymbol s}}
\def\hs{\widehat{ s}}
\def\cS{{\cal S}}
\def\bt{{\boldsymbol t}}
\def\bT{{\boldsymbol T}}
\def\bu{{\boldsymbol u}}
\def\hbu{\widehat{\boldsymbol u}}
\def\bU{{\boldsymbol U}}
\def\tu{\tilde{u}}
\def\bv{{\boldsymbol v}}
\def\bV{{\boldsymbol V}}
\def\bw{{\boldsymbol w}}
\def\bW{{\boldsymbol W}}
\def\tw{\tilde{w}}
\def\bx{{\boldsymbol{x}}}
\def\tx{\widetilde{x}}
\def\tbx{\widetilde{\boldsymbol{x}}}
\def\bX{{\boldsymbol X}}
\def\cX{{\cal X}}
\def\tX{\widetilde{X}}
\def\ty{\tilde{y}}
\def\by{{\boldsymbol y}}
\def\bY{{\boldsymbol Y}}
\def\tby{\tilde{\boldsymbol y}}
\def\tY{\widetilde{Y}}
\def\hy{\hat{y}}
\def\byhat{{\hat {\boldsymbol y}}}
\def\tY{\widetilde{Y}}
\def\cY{{\cal Y}}
\def\Ybar{\overline{Y}}
\def\bz{{\boldsymbol z}}
\def\bZ{{\boldsymbol Z}}
\def\cZ{{\cal Z}}
\def\tz{\tilde{z}}
\def\tZ{\widetilde{Z}}
\def\brho{{\boldsymbol{\rho}}}
\def\bzero{{\boldsymbol 0}}
\def\eps{\epsilon}
\def\veps{\varepsilon}
\def\bveps{\boldsymbol{varepsilon}}
\def\tveps{\widetilde{\varepsilon}}
\def\tbveps{\widetilde{\boldsymbol{\varepsilon}}}
\def\Ghat{\widehat{G}}
\def\argmax{\mathop{\rm argmax}}
\def\argmin{\mathop{\rm argmin}}
\def\real{\mathop{{\rm I}\kern-.2em\hbox{\rm R}}\nolimits}
\def\diag{\mbox{diag}}

\def\sgn{\hbox{sgn}}
\def\Var{\hbox{Var}}
\def\Cov{\hbox{Cov}}
\def\Rem{\hbox{Rem}}

\def\whbeta{\widehat{\beta}}
\def\hbeta{\hat{\beta}}
\def\whtheta{\widehat{\theta}}
\def\htheta{\hat{\theta}}
\def\whF{\widehat{F}}
\def\hF{\hat{F}}
\def\Dnu{\Delta_{\nu}}
\def\median{\mbox{median}}
\def\sign{\mbox{sign}}
\def\trace{\mbox{trace}}

\def\bone{{\boldsymbol 1}}
\def\bzero{{\boldsymbol 0}}
\def\balpha{\boldsymbol \alpha}
\def\bkappa{\boldsymbol \kappa}
\def\btheta{\boldsymbol \theta}
\def\bbeta{\boldsymbol \beta}
\def\bgamma{\boldsymbol \gamma}
\def\hbbeta{\hat{\boldsymbol \beta}}
\def\hsbbetan{\hat{\boldsymbol \beta_n^*}}
\def\tbeta{\tilde{\beta}}
\def\tbbeta{\tilde{\boldsymbol \beta}}
\def\bdelta{\boldsymbol \delta}
\def\bata{\boldsymbol \eta}
\def\gam{\gamma}
\def\lam{\lambda}
\def\blam{\boldsymbol \lambda}
\def\hmu{\widehat{\mu}}
\def\bmu{\boldsymbol \mu}
\def\bnu{\boldsymbol \nu}
\def\hphi{\widehat{\phi}}
\def\drho{\dot{\rho}}
\def\hsigma{\widehat{\sigma}}
\def\ttheta{\widetilde{\theta}}
\def\hbtheta{\widehat{\boldsymbol \theta}}
\def\bveps{\boldsymbol \varepsilon}
\def\tbveps{{\tilde\bveps}}
\def\bxi{\boldsymbol \xi}
\def\txi{\tilde{\xi}}
\def\tzeta{\tilde{\zeta}}

\begin{abstract}
{ In genome-wide association studies (GWAS), penalization is an important approach for identifying genetic markers associated with trait while mixed model is successful in accounting for a complicated dependence structure among samples. Therefore, penalized linear mixed model is a tool that combines the advantages of penalization approach and linear mixed model. In this study, a GWAS with multiple highly correlated traits is analyzed. For GWAS with multiple quantitative traits that are highly correlated, the analysis using traits marginally inevitably lose some essential information among multiple traits. We propose a penalized-MTMM, a penalized multivariate linear mixed model that allows both the within-trait and between-trait variance components simultaneously for multiple traits. The proposed penalized-MTMM estimates variance components using an AI-REML method and conducts variable selection and point estimation simultaneously using group MCP and sparse group MCP. Best linear unbiased predictor (BLUP) is used to find predictive values and the Pearson's correlations between predictive values and their corresponding observations are used to evaluate prediction performance. Both prediction and selection performance of the proposed approach and its comparison with the uni-trait penalized-LMM are evaluated through simulation studies. We apply the proposed approach to a GWAS data from Genetic Analysis Workshop (GAW) 18.  
}
\end{abstract}
\let\thefootnote\relax\footnotetext{$^{\dagger}$Co-first authors. $^*$ To whom correspondence should be addressed.}

\section{Introduction}
Genome-wide association studies (GWAS) help us better understand the genetic basis of many complex traits~\citep{gwas_rev_08}. A better understanding of the relationship between phenotypic trait and genetic variation for these quantitative and complex traits will yield insights that are essential to predict disease risk and develop personalized therapeutic treatments for human-beings. At the beginning stage of GWAS, researchers mainly focused on a single trait analysis~\citep{burton2007genome}. Although GWAS have identified some of the genetic risk variants~\citep{visscher2012five}, those identified variants can only explain a small fraction of phenotypic variance, which is known as the ``missing heritability'' problem~\citep{manolio2009finding}. Recent analysis suggested that a substantial proportion of heritability was not missing but hidden in the common variants with small or moderate effects ~\citep{yang2010common,makowsky2011beyond}.

On one hand, these results suggest that recruiting a large sample size will help to identify genetic risk variants but it could be very expensive. On the other hand, researchers start to be interested in simultaneously analyzing multiple correlated traits recently to improve the statistical power~\citep{korte2012mixed}.  This is because the correlated traits may share common genetic factors, which is known as pleiotropy \citep{sivakumaran2011abundant}. For example, a ``pleiotropic enrichment'' method was applied to analyze the GWAS data sets of schizophrenia and cardiovascular disease. The power to detect schizophrenia-associated common
variants was shown to be improved by exploiting the pleiotropy between these two phenotypes. More recently, a study on genome-wide SNP data for five psychiatric disorders in
33,332 cases and 27,888 controls identified four significant loci ($P < 5\times 10^{-8}$) affecting
multiple disorders~\citep{Consortium2013identification}. It is expected that successfully taking account for the pleiotropy structure will be helpful for identification of risk variants.

In this study, a GWAS from GAW 18 with multiple traits that are highly correlated is analyzed. This type of data exposes an opportunity to integratively analyze multiple traits from a GWAS. In this paper, we focus on GAW 18 data with two highly correlated traits -- systolic blood pressure (SBP) and diastolic blood pressure (DBP). We propose a unified framework to simultaneously analyze multiple traits, in which we introduce a variance component to account for sample relatedness or the confounding effects of population stratification, and introduce some sparse penalties to detect risk variants.  Our approach bridges the advantages of multi-trait linear mixed models with penalized regression techniques. For the choice of sparse penalties, we have two types of models --- homogeneity and heterogeneity models. Homogeneity model assumes that all trait-associated markers/variants are consistent across all traits, while heterogeneity model assumes that a marker/variant may be associated with some traits but not others. Depending on the assumption of homogeneity and heterogeneity, group MCP and sparse group MCP can be used to conduct variable selection.

The rest of the article is organized as follows. In Section 2, we show the data structure and review variance components model in genetics. The estimation of variance components, predictions, penalized selection and method to collapse SNPs are described in Section 3. Numerical studies, including simulation in Section 4 and data analysis in Section 5, are conducted to investigate finite sample performance. The article concludes with discussion in Section 6.

\section{Data and Model}
\label{data_description}
\subsection{GAW 18 Data}
The genetic analysis workshop (GAW) 18 provides the type 2 diabetes genetic exploration by next--generation sequencing in ethnic samples (T2D--GENES) consortium data set that consists of 1,043 individuals from 20 Mexican American pedigrees enriched for type 2 diabetes from San Antonio, TX. The study included subjects in two different groups, including the San Antonio family heart study (SAFHS) and the San Antonio family diabetes/gallbladder study (SAFDGS), which are together referred to as the San Antonio family studies (SAFS). Whole genome sequence is being performed commercially at Complete Genomics, Inc and the GAW 18 data set is based on the sequence data for the first 483 T2D--GENES. GWAS data for 472,049 SNPs on odd numbered autosomes are provided for these 959 family members (464 directly sequenced and the rest imputed~\citep{howie2012fast}). A variety of different phenotypic traits were measured at examination, e.g. systolic blood pressure (SBP), diastolic blood pressure (DBP). Clearly, SBP and DBP are highly correlated traits.
GAW 18 data set brings a good opportunity to develop statistical models to handle multiple correlated traits in the pedigree-based samples. We aim at identifying risk variants while accounting for the correlation among multiple traits and the relatedness among the samples.

\subsection{Variance Components Model in Genetics}
Recently, mixed model has been extensively studied for correcting the genetic relatedness in association mapping in genome-wide association studies (GWAS). The genetic relatedness from population mixture and inbred strains can cause the problem of inflated false positive rates. However, most of the existing methods fail to consider the mixed models with multiple traits. Denote that in a GWAS, we have $n$ subjects and $p$ genes of genetic scores with $m$ traits. Assume that we have two traits---trait 1 and trait 2. Note that it can be relaxed to more than two traits. For each trait, the relatedness matrix ($K$) can be used to describe the genetic relatedness. For multiple traits, we vectorize the multiple traits. When it comes to the analysis of multiple traits, a natural extension is to use $\mathrm{diag}(K,K)$. Here, we still miss a variance component describing the relatedness between/among multiple traits. \cite{bio_multi_12} used the covariance components among random effects across multiple traits to describe this relatedness. We go further on this direction including covariance components among residuals across multiple traits. The variance covariance matrix for vectorized two traits is given:
\begin{equation}
\label{var_covar}
VC=
\begin{pmatrix}
 K\sigma_{g^{(1)}}^2 + \mathrm{I}_n \sigma_{e^{(1)}}^2 & K\sigma_{g^{(12)}} + \mathrm{I}_n \sigma_{e^{(12)}} \\
 K\sigma_{g^{(12)}} + \mathrm{I}_n \sigma_{e^{(12)}} & K\sigma_{g^{(2)}}^2 + \mathrm{I}_n \sigma_{e^{(2)}}^2
\end{pmatrix},
\end{equation}
where $\sigma_{g^{(1)}}^2$ and $\sigma_{g^{(2)}}^2$ are variance components for random effects on trait 1 and trait 2, $\sigma_{g{^{(12)}}}$ is the covariance of random effects between trait 1 and trait 2, $\sigma_{e^{(1)}}^2$ and $\sigma_{e^{(2)}}^2$ are variance components for residuals on trait 1 and trait 2, and $\sigma_{e^{(12)}}$ is the covariance of residuals between trait 1 and trait 2. 
We implement ``Average information - restricted maximal likelihood'' method (AI-REML)~\citep{AIREML95} to estimate variance components. With the variance components fixed, we may implement penalization methods to conduct variable selection and point estimation simultaneously. More details are discussed in Section 3.

\section{Variance Components and Penalized Regression}
Let us first consider the linear mixed model (LMM) which is widely used in single-trait analysis ~\citep{mixed_10,var_10,fastLMM11,gemma_12,rakitsch2013lasso} and then extend it to handle multiple correlated traits. Let $n$ be the sample size, the LMM can be written as:
\begin{equation}\label{lmm1}
\begin{aligned}
    &y_o=Wv + X b + g + e,\\
    &g \sim \mathrm{N}(0,\sigma^2_g \mathrm{K}),\\
    &e \sim \mathrm{N}(0,\sigma^2_e \mathrm{I}),
\end{aligned}
\end{equation}
where $y_o\in \mathbb{R}^{n\times 1}$ is the response vector representing the trait, $W\in \mathbb{R}^{n\times q}$ is the matrix of covariates (fixed effect) including the intercept and other covariates such as age and gender, $b$ is the vector for regression coefficients of the covariates, $X \in \mathbb{R}^{n\times p}$ is the genotype matrix and $b$ is the vector for the effect sizes of $p$ SNPs (fixed effects), $g$ is the random effect from $\mathrm{N}(\mathbf{0},\sigma^2_g\mathrm{K})$, and $e$ is the residual error with variance $\sigma^2_e$. Here the covariance matrix $\mathrm{K}$ is the genetic relatedness matrix which describes the pedigree structure among the individuals and $\sigma^2_g$ is the variance component of $g$. The covariance matrix $\mathrm{K}$ can be constructed according to the known pedigree information or estimated from genome-wide SNP information. This model can be interpreted as follows: The random effect $g$ can be considered as a global average of signals from the genetic background
and the shared environmental influence and we call it ``average polygenic effect''. For those SNPs with large effects
which are different from the genetic background, they are put into the design matrix $X$ and considered as fixed effects. In this way, the markers with larger effects can be
treated locally.

\subsection{Computation of Variance Components}
Now we extend single-trait model (\ref{lmm1}) to a multiple-trait model. Let $y_o$ be an $n\times m$ response matrix with each row representing subject and each column representing a trait. Let ${\bf g}(=(g^{(1)},\dots,g^{(m)}))$ and ${\bf e}(=(e^{(1)},\dots,e^{(m)}))$ be $n\times m$ matrix of unobserved polygenic and random residual effects, respectively. Denote $W$ be $n\times q$ non-genetic covariates and $X$ be $n\times p$ genetic scores of candidate genes. Correspondingly, we denote $V(=(v^{(1)},\dots,v^{(m)}))$ and $B(=(b^{(1)},\dots,b^{(m)}))$ the corresponding $q\times m$ coefficient matrix for $q$ non-genetic covariates and $p\times m$ coefficient matrix for $p$ genetic scores in $m$ traits. We denote $y_o^{(l)}$ the $l$th trait and further denote that $v^{(l)}$, $b^{(l)}$, $g^{(l)}$ and $e^{(l)}$ are the corresponding vectors of coefficient for non-genetic effects, genetic effects, average polygenic effects and vector of random residual, respectively, for $l(=1,\dots,m)$. First, consider linear mixed model for $l$th trait:
\begin{equation*}
\label{model_mar}
 y_o^{(l)} = Wv^{(l)} + Xb^{(l)} + g^{(l)} + e^{(l)},
\end{equation*}
where $g^{(l)}\sim\mathrm{N}(\boldsymbol{0},\mathrm{K}\sigma_{g^{(l)}}^2)$ and $e^{(l)}\sim\mathrm{N}(\boldsymbol{0},\mathrm{I}_n\sigma_{e^{(l)}}^2)$. $\sigma_{g^{(l)}}^2$ and $\sigma_{e^{(l)}}^2$ are variance components describing relations among subjects. In order to account for the genetic correlation and residual correlation for multiple traits,  we introduce $\sigma_{g^{(l,k)}}$ and $\sigma_{e^{(l,k)}}$ to describe the covariance of average polygenic effects and residual for $l$th and $k$th trait, respectively. Consider the multivariate linear mixed model:
\begin{equation}
\label{model}
 \mathbf{y}_o = WV + XB + \mathbf{g} +\mathbf{ e},
\end{equation}
where we assume that
\begin{enumerate}
 \item $\mathrm{vech}({\bf e})\sim \mathrm{N}(\boldsymbol{0}, \mathrm{I}_n \otimes\Sigma_e)$ where $\Sigma_e$ is a $m\times m$ matrix describing covariance structure among multiple traits.
 \item $\mathrm{vech}({\bf g})\sim \mathrm{N}(\boldsymbol{0},\mathrm{K} \otimes \Sigma_g)$ where $\Sigma_g$ is $m\times m$ matrix describing covariance structure among multiple traits. $\mathrm{K}$ is an $n\times n$ genetic relatedness matrix (twice of the kinship matrix) and it can be calculated using genome-wide genetic markers \citep{yang2010common} or directly obtained from unknown pedigree information.
\end{enumerate}

%

There is a difficulty in applying the model when $q+p+m(m+1)>n$, when the number of
parameters exceeds the number of samples ($d$ is the number of covariates, $p$ is the
number of SNPs treated as fixed effects, $m(m+1)$ is the number of variance components). In
order to overcome this difficulty, we introduce sparse constraints on $b$ to perform variable selection, such that model (\ref{model}) is well defined.
To incorporate the feature of homogeneity and heterogeneity structure among traits, we propose to use group MCP and sparse group MCP. We call this approach as ``Penalized Multi-Trait Mixed Models (Penalized-MTMM)''.

For simplicity, here we only consider two traits ($m=2$) but the framework for more than two traits remains the same. For $m = 2$, we have $\Sigma_e=\begin{pmatrix}
 \sigma_{e^{(1)}}^2 & \sigma_{e^{(12)}} \\
 \sigma_{e^{(12)}} & \sigma_{e^{(2)}}^2
\end{pmatrix}$ and $\Sigma_g=\begin{pmatrix}
 \sigma_{g^{(1)}}^2 & \sigma_{g^{(12)}} \\
 \sigma_{g^{(12)}} & \sigma_{g^{(2)}}^2
\end{pmatrix}$.  Denote that $S=\mathrm{diag}(W,W)$, $v=\mathrm{vech}(V)$, $T=\mathrm{diag}(X,X)$, $b=\mathrm{vech}(B)$, $y=\mathrm{vech}({\bf y_o})$, $g=\mathrm{vech}({\bf g})$ and $e=\mathrm{vech}({\bf e})$. Model (\ref{model}) becomes
\begin{equation}\label{mtmm1}
\begin{aligned}
    &y= Sv+Tb+g+e,\\
    &g \sim \mathrm{N}(\boldsymbol{0},\mathrm{K} \otimes \Sigma_g),\\
    &e \sim \mathrm{N}(\boldsymbol{0}, \mathrm{I}_n \otimes\Sigma_e),
\end{aligned}
\end{equation}
Integrating out $g$ and $e$ and we have $y\sim \mathrm{N}(Sv+Tb,\mathrm{K} \otimes \Sigma_g + \mathrm{I}_n \otimes\Sigma_e)$. The log-likelihood can be analytically written as
\begin{equation}\label{}
  L(v,b,\Sigma_g,\Sigma_e) = -\frac{1}{2}\left[ 2n \log(2\pi) + \log(|H|) + (y-Sv-Tb)^T H^{(-1)}(y-Sv-Tb)\right].
\end{equation}
where $H = \mathrm{K} \otimes \Sigma_g + \mathrm{I}_n \otimes\Sigma_e$. Now we introduce sparse penalties on the coefficient $b$ and the penalized log-likelihood can be written as
\begin{equation}\label{pen-loglike-MTMM}
  L(v,b,\Sigma_g,\Sigma_e) = -\frac{1}{2}\left[ 2n \log(2\pi) + \log(|H|) + (y-Sv-Tb)^T H^{(-1)}(y-Sv-Tb)\right] -  P_\lambda (b).
\end{equation}
where $\lambda$ is the regularization parameter.


Clearly, when $b$ is fixed, the optimization of penalized log-likelihood function (\ref{pen-loglike-MTMM}) can be solved by the standard ``Average information - restricted maximal likelihood'' method (AI-REML) ~\citep{AIREML95}. When ($v,\Sigma_g,\Sigma_e$) are all known, we will show that maximization of penalized log-likelihood becomes a penalized least square problem. We will carefully discuss different penalties in next section.

\subsection{Computation of Penalized-MTMM}
We begin with $b=0$ and solve (\ref{pen-loglike-MTMM}) by AI-REML. After obtaining ($v,\Sigma_g,\Sigma_e$), we can transform the log-likelihood function (\ref{pen-loglike-MTMM}) to penalized least square problem as follows. Let $\hat{H}$ and $\hat{v}$ be the estimate of $H$ and $v$, given by AI-REML, respectively. Denote that $\tilde{y}={\hat{H}}^{-1/2}(y-S\hat{v})$, and $\tilde{T}={\hat{H}}^{-1/2}T$. Ignoring some constants, the unpenalized log-likelihood (\ref{pen-loglike-MTMM}) becomes ($L(b)$=)$\lVert\tilde{y}-\tilde{T}b\rVert^2/2$. Hence, the maximization of the regularized penalized log-likelihood (\ref{pen-loglike-MTMM}) is equivalent to the following optimization problem
\begin{equation}
 \min_{b} \frac{1}{n}L(b)+ \mathrm{P}_{\lambda}(b),
\end{equation}
where $\mathrm{P}_{\lambda}(b)$ is a penalty function on the effects of genetic variants.

Similar to integrative analysis~\citep{Integrative11}, we can assume homogeneous or heterogeneous structure across multiple traits. Homogeneity model assumes that both traits share the same set of trait-associated covariates while heterogeneity model assumes that a covariate can be associated with some of traits but not others. For homogeneity model, group MCP has been demonstrated to conduct variable selection effectively, while sparse group MCP can be used to conduct variable selection between- and within-groups for heterogeneity model. Here, we choose to use minimax concave penalization (MCP) as basic penalty for the variant selection, since comparing with its alternative, e.g. Lasso~\citep{Tib96} and smooth clipped absolute deviation (SCAD)~\citep{FL01}, MCP belongs to the family of quadratic spline penalties and leads to oracle selection results requiring weaker conditions~\citep{ZCH10}. We refer to \cite{ZCH10} and \cite{MFH09} for detailed discussion.

The MCP is defined as
\[
\rho(t;\lambda, \gamma) = \lambda_1 \int_0^{|t|} (1-x/(\gamma
\lambda))_{+} dx.
\]
Here $\lambda$ is a penalty parameter, $\gamma$ is a regularization parameter that controls the concavity of $\rho$ and $x_+= x 1_{\{x \ge 0\}}$. The MCP can be easily understood by considering its derivative, which is
\begin{equation}
\dot{\rho}(t;\lambda,\gamma) = \lambda_1 \big(1-|t|/(\gamma
\lambda)\big)_{+}\mathrm{sgn}(t),  \nonumber
\end{equation}
where $\mathrm{sgn}(t)=-1, 0,$ or $1$ if $t < 0, =0$, or $> 0$, respectively. As $|t|$ increases from 0, MCP begins by applying the same rate of penalization as Lasso, but continuously relaxes that penalization until $|t| > \gamma\lambda$, a condition under which the rate of penalization drops to 0. It provides a continuum of penalties where the Lasso penalty corresponds to $ \gamma = \infty$ and the hard-thresholding penalty corresponds to $\gamma \rightarrow 1+$. We note that other penalties, such as Lasso or SCAD, can also be used to replace MCP. We choose MCP because it possesses all the desirable properties of a penalty function and is computationally simple~\citep{MFH09,ZCH10}.

\subsubsection{Group MCP}
For the homogeneity model, group penalization methods can be implemented, e.g. group Lasso, group bridge, group MCP. Here, we choose to use group MCP~\citep{HWM11} since comparing with its alternative, it possesses oracle properties with less conditions. 
We have $b=\mathrm{vech}(B)$ where $B_j$ is the $j$th row of $B$ corresponding to the regression coefficients of the $j$th gene on multiple traits, $\lVert B_j\rVert_{\Sigma_j}=(B_j^\prime\Sigma_jB_j)^{1/2}$ and $\Sigma_j=\tilde{T}_j^\prime\tilde{T}_j/n$ is the empirical covariance matrix for the $j$th group. We can write $\Sigma_j=\mathit{R}_j^\prime{\mathit{R}}_j$ for an $m\times m$ upper triangular matrix $\mathit{R}_j$ with positive diagonal entries via the Cholesky decomposition. Let $V_j=\tilde{T}_j\mathit{R}_j^{-1}$ and $\beta_j=\mathit{R}_jB_j$. With the transformation, the penalty function of group MCP is $\mathrm{P}_{\lambda}(\bbeta)=\sum_{j=1}^p \rho\left(\lVert \beta_j\rVert;\sqrt{m}\lambda,\gamma\right)$ and the objective function corresponding to group MCP~\citep{HWM11} can be written as:
\begin{equation}
\label{gMCP_obj}
 L_{GM}(\bbeta,\lambda)=
 \frac{1}{2n}\lVert\tilde{y}-\sum_{j=1}^p V_j\beta_j\rVert^2 +  \mathrm{P}_{\lambda}(\bbeta),
\end{equation}
where $\bbeta=(\beta_1^{\prime},\ldots, \beta_p^{\prime})^{\prime}$.

The rationale behind the group MCP can also be understood by its univariate solution with the $j$th group. Consider the linear regression of $y$ upon $x_j$ (covariates on the $j$th group), with unpenalized least squares solution $z_j=n^{-1}x_j^{\prime}y$ (recall that $x_j$ has been standardized so that $x_j^{\prime}x_j/n = \mathrm{I}_n$). For this linear regression problem, the group MCP estimator has the following closed form:
\begin{equation*}
 \hat{\beta_j}=\begin{cases}
 \frac{\gamma}{\gamma-1} \mathrm{S}_1(z_j,\sqrt{m}\lambda), & \text{if }\| z_j\|_2\le a\sqrt{m}\lambda \\
 z_j, & \text{if }\|z_j\|_2> \gamma\sqrt{m}\lambda
\end{cases}, \label{gMCP_solution}
\end{equation*}
where $\mathrm{S}_1(z, \lambda)=\left(1-\lambda/\lVert z\rVert_2 \right)_+ z$. Then one can implement group coordinate descent (GCD) algorithm to solve for the optimizer of objective function (\ref{gMCP_obj})~\citep{HWM11}.

\subsubsection{Sparse Group MCP}
For heterogeneity structure among traits, sparse group MCP can be applied to conduct variable selection. 
We orthogonalize covariates within groups in the same fashion as the group MCP. Denote $\boldsymbol{\lambda}=(\lambda_1,\lambda_2)$. Then, the penalty function of sparse group MCP is $\mathrm{P}_{\boldsymbol{\lambda}}(\bbeta)=\sum_{j=1}^p \rho\left(\lVert \beta_j\rVert;\sqrt{m}\lambda_1,\gamma\right)+\sum_{j=1}^p\sum_{k=1}^m \rho\left(\lvert \beta_{jk}\rvert;\lambda_2,\gamma\right)$ and the objective function corresponding to sparse group MCP~\citep{LHM_sparse_SIM_13} can be written as:
\begin{equation}
\label{sparse_gMCP_obj}
 L_{SGM}(\bbeta,\lambda)=
 \frac{1}{2n}\lVert\tilde{y}-\sum_{j=1}^p V_j\beta_j\rVert^2 +  \mathrm{P}_{\boldsymbol{\lambda}}(\bbeta).
\end{equation}

Similar to~\cite{ZL_LLA_08,BH09}, one can use local linear approximation (LLA) for the penalty function and iteratively solve the problem using the optimizer in~\cite{Sparse10}. In \cite{LHM_sparse_SIM_13}, they used a two-step strategy to solve for the optimizer of objective function (\ref{sparse_gMCP_obj}). In this way, GCD algorithm can be implemented to solve (\ref{sparse_gMCP_obj}) instead of solving objective function with LLA penalty function. \cite{BH10} argued that GCD algorithm, alternative to LLA can be implemented with more efficiency. Consider univariate group solution on the linear regression of $y$ upon $x_j$ (covariates on the $j$th group), with unpenalized least squares solution $z_j=n^{-1}x_j^{\prime}y$ (recall that $x_j$ has been standardized so that $x_j^{\prime}x_j/n = \mathrm{I}_n$).

By setting first order derivative of objective function be zero, we have:
\begin{equation}
 -z_j + g(\beta_j)\beta_j+\mathbf{t}=0, \label{s_mcp}
\end{equation}
where $z_j=(z_j^1,\dots,z_j^M)^{\prime}$, $g(\beta_j)=\left(1+\frac{1}{||\beta_j||_2}\right)\begin{cases}
 \sqrt{m}\lambda_1-\frac{\|\beta_j\|_2}{\gamma}, & \text{if }\|\beta_j\|_2\le \gamma\sqrt{m}\lambda_1  \\
 0, & \text{if }\|\beta_j\|_2> \gamma\sqrt{m}\lambda_1
\end{cases}.$
Denote $z_j^k$ as the $k$th element of $z_j$. First, fix $g(\beta_j)$ at the current estimate $\tilde\beta_j$, we use $g$ short for $g(\tilde\beta_j)$. The $k$th element in equation~(\ref{s_mcp}) can be rewritten as:
\begin{equation}
 -\frac{z_{jk}}{g} + \beta_{jk}+\mathrm{sgn}(\beta_{jk})\begin{cases}
\frac{\lambda_2}{g}-\frac{|\beta_{jk}|}{\gamma g}, & \text{if }|\beta_{jk}|\le \gamma\lambda_2 \\
 0, & \text{if }|\beta_{jk}|> \gamma\lambda_2 \label{res_mcp}
\end{cases}  = 0.
\end{equation}
The solution to equation~(\ref{res_mcp}) is $$\widehat{g\beta_{jk}}= \begin{cases} \frac{\mathrm{S}_2(z_{jk},\lambda_2)}{1-\frac{1}{\gamma g}},& \text{if }|z_{jk}|\le \gamma\lambda_2 g\\
z_{jk}, & \text{if }|z_{jk}|> \gamma\lambda_2 g.
\end{cases}$$
Here $\mathrm{S}_2(z, \lambda)=\mathrm{sgn}(z)(|z|-\lambda)_+$.
For $k=1,\dots,m$, set $u_k = \widehat{g\beta_{jk}}$ and $\mathbf{u}=(u_1,\dots,u_{m})^{\prime}$. Taking $\mathbf{u}$ back into its definition,
\begin{equation}
\beta_j+\frac{\beta_j}
{\|\beta_j\|_2}\begin{cases}
\sqrt{m}\lambda_1-\frac{\|\beta_j\|_2}{a}, & \text{if }\|\beta_j\|_2\le a\sqrt{m}\lambda_1  \\
0, & \text{if }\|\beta_j\|_2> a\sqrt{m}\lambda_1
\end{cases} =  \mathbf{u}.  \label{2nd_sparse}
\end{equation}
Expression~(\ref{2nd_sparse}) can be solved in a similar manner as with the gMCP, leading to
\begin{equation}
 \hat{\beta}_j=\begin{cases}
 \frac{a}{a-1} \mathrm{S}_1(\mathbf{u},\sqrt{m}\lambda_1), & \text{if }\| \mathbf{u}\|_2\le a\sqrt{m}\lambda_1 \\
 \mathbf{u}, & \text{if }\|\mathbf{u}\|_2> a\sqrt{m}\lambda_1
\end{cases}. \label{solution}
\end{equation}

To optimize the group MCP or sparse group MCP objective function, group coordinate descent algorithm (GCD) can be implemented. \cite{BH10} explored coordinate descent algorithms (CDA) for nonconvex penalized regression, including MCP and SCAD. The extension of CDA to group level is natural, their details can be found in \cite{HWM11,LHM_sparse_SIM_13}.

\subsubsection{Choice of tuning parameter}
With MCP, there is one tuning parameter $\lambda$ and one regularization parameter $\gamma$. Generally speaking, smaller values of $\gamma$ are better at retaining the unbiasedness of the MCP penalty for large coefficients, but they also have the risk of creating objective functions with a nonconvexity problem that are difficult to optimize and yield solutions that are discontinuous with respect to $\lambda$. It is therefore advisable to choose a $\gamma$ value that is big enough to avoid this problem but not too big. Simulation studies in~\citet{BH10} and~\cite{Int_surv12} show that $\gamma=3$ is a reasonable choice for group MCP and $\gamma=6$ is a reasonable choice for sparse group MCP, respectively. For group MCP, we search for tuning parameters $\lambda$ using $V$-fold cross validation ($V=5$ in our numerical study). For sparse group MCP, we fix the ratio of $\lambda_1$ and $\lambda_2$ to be 1. Then $\lambda_1$ and $\lambda_2$ can be searched through $V$-fold cross validation. It is expected that tuning parameter cannot go down to very small values which correspond to regions not locally convex. The cross validation criteria over non-locally convex regions may not be monotone. More details regarding the choice of tuning parameter for group MCP and sparse group MCP can be found in~\citep{HWM11} and~\cite{Int_surv12}, respectively.

\subsection{Genetic Scores on Collapsed SNPs}
\label{collapse_snp}
Recent GWAS have shown that common variants can only account for small proportion of heritability. Among all potential explanations to this missing heritability, the large number of variants of small effects and rare variants (possibly with large effects) can be partially remedied using a weighted-sum method~\citep{wsum_09}. We group SNPs at gene level using this weighted-sum method. This process puts the analysis for genetic markers at gene level and is capable of dealing with rare variants together with common variants. The proposed approach can be easily implemented in GWAS with common variants only or longitudinal measurements on traits with independent samples.

\subsection{Trait Prediction}
\label{prediction}
Given a training sample of genetic variants and traits, we can predict the unobserved traits using a testing sample. According to best linear unbiased predictor (BLUP), the predictive value of $l$th trait $y_p^{(l)}$ is given by
\begin{equation}
\label{blup}
  y_p^{(l)} =W\hat{v}^{(l)}+X\hat{b}^{(l)}+K_{tt}{\hat{H}^{(l)}}^{-1}(y_t^{(l)}-W\hat{v}^{(l)}-X\hat{b}^{(l)})
\end{equation}
where $y_t^{(l)}$ is the $l$th trait of the training set, $K_{tt}$ is the covariance matrix between the training sample and the testing sample and $\hat{H}^{(l)}$ is matrix of variance components corresponding to $l$th trait. To evaluate estimates from the penalized-MTMM, we first extract variance components and genetic effects corresponding to each trait. Then, we marginally evaluate the prediction on each trait and calculate the Pearson's correlation between the predictive values and their corresponding observations. This procedure puts the comparison between penalized-MTMM and uni-trait penalized-LMM methods on the same page. 

\section{Simulation Study}
We conduct simulation to better gauge performance of the proposed methods. The genotype data is excerpted from a T2D--GENES study with twenty pedigree families (Section~\ref{data_description}). We consider six scenarios of correlations. 
For all scenarios, we set $n=400$, $p=5000$ and $m=2$. We consider two traits in this study. The covariance among residual ($\Sigma_e$) and random effects ($\Sigma_d$) in six scenarios are listed in Table~\ref{Tab:01}. Scenario 1--3 represent the cases that $\Sigma_e$ and $\Sigma_d$ is proportional with weak, moderate and strong correlation while scenario 4--6 represent the cases that $\Sigma_e$ and $\Sigma_d$ is not proportional with multiple combination on $\Sigma_e$ and $\Sigma_d$. We also consider homogeneity and heterogeneity structure between two traits in this simulation study. In homogeneity structure, the index of important variants are (1--5, 16--20) for both traits while the index of important variants are (1--5, 16--20) for trait 1 and (1--5, 21--25) for trait 2 in heterogeneity structure. 

\begin{figure}[ht]
\centering\subfigure[Scenario 1]{\label{case1}\includegraphics[scale=0.3]{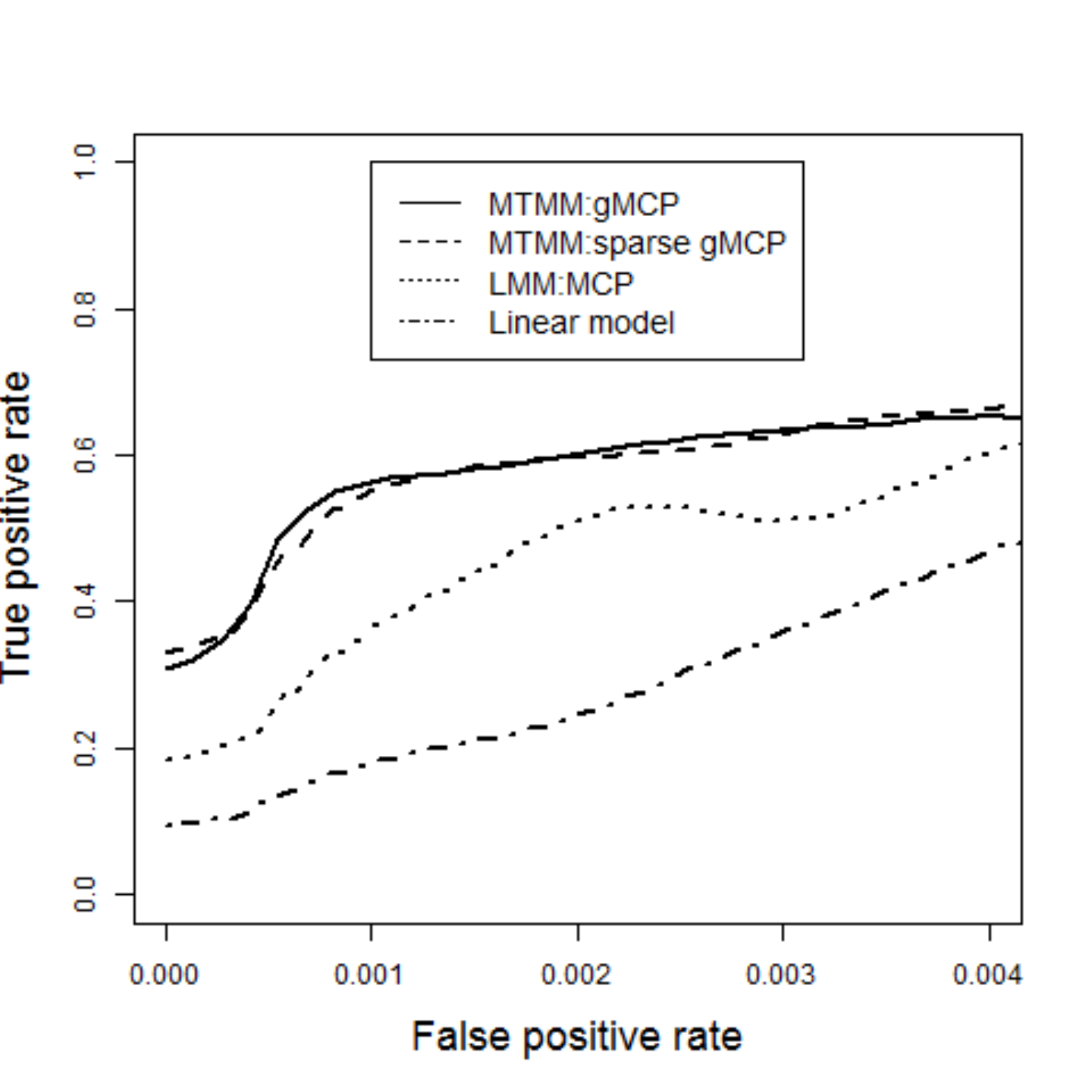}}
\centering\subfigure[Scenario 2]{\label{case2}\includegraphics[scale=0.3]{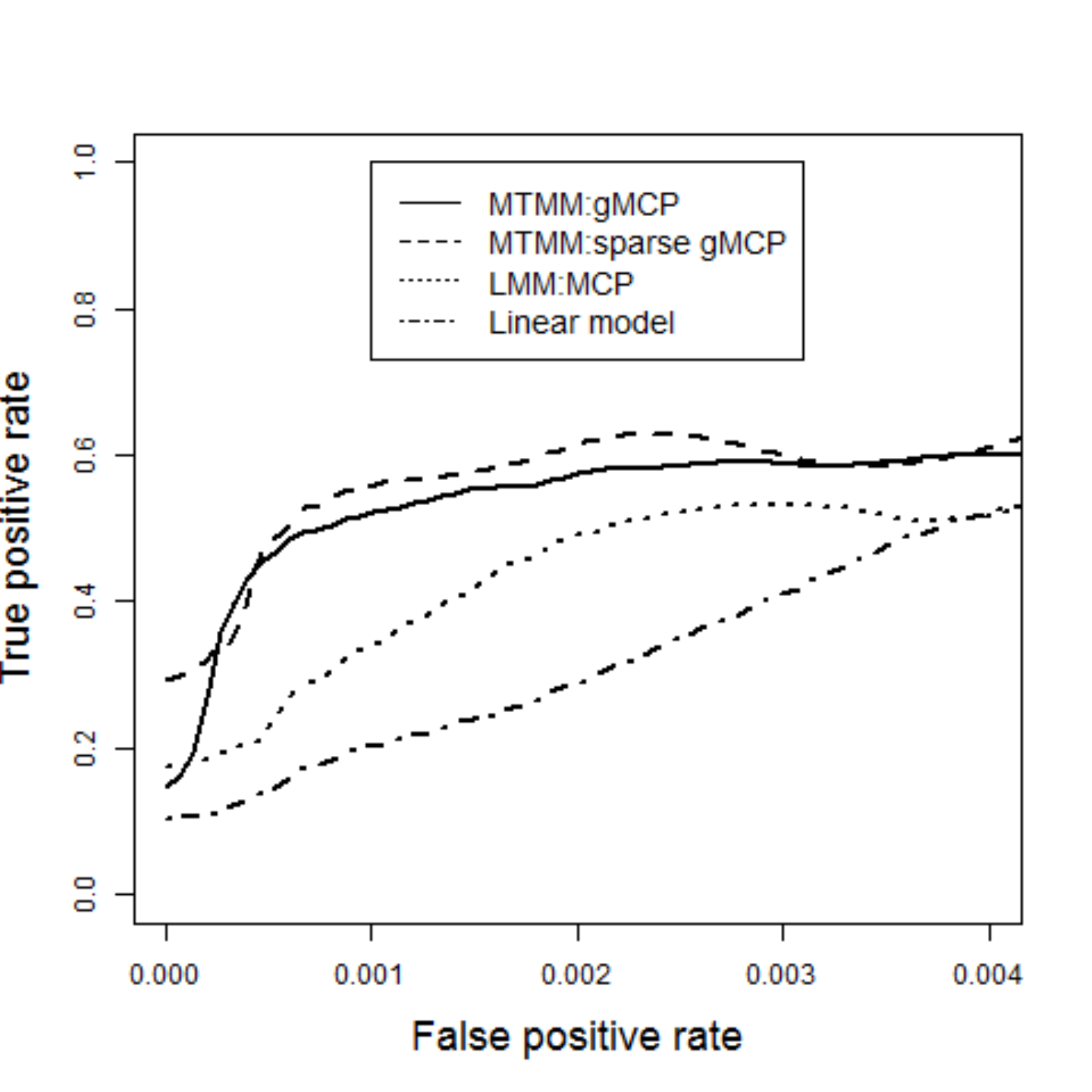}}
\centering\subfigure[Scenario 3]{\label{case3}\includegraphics[scale=0.3]{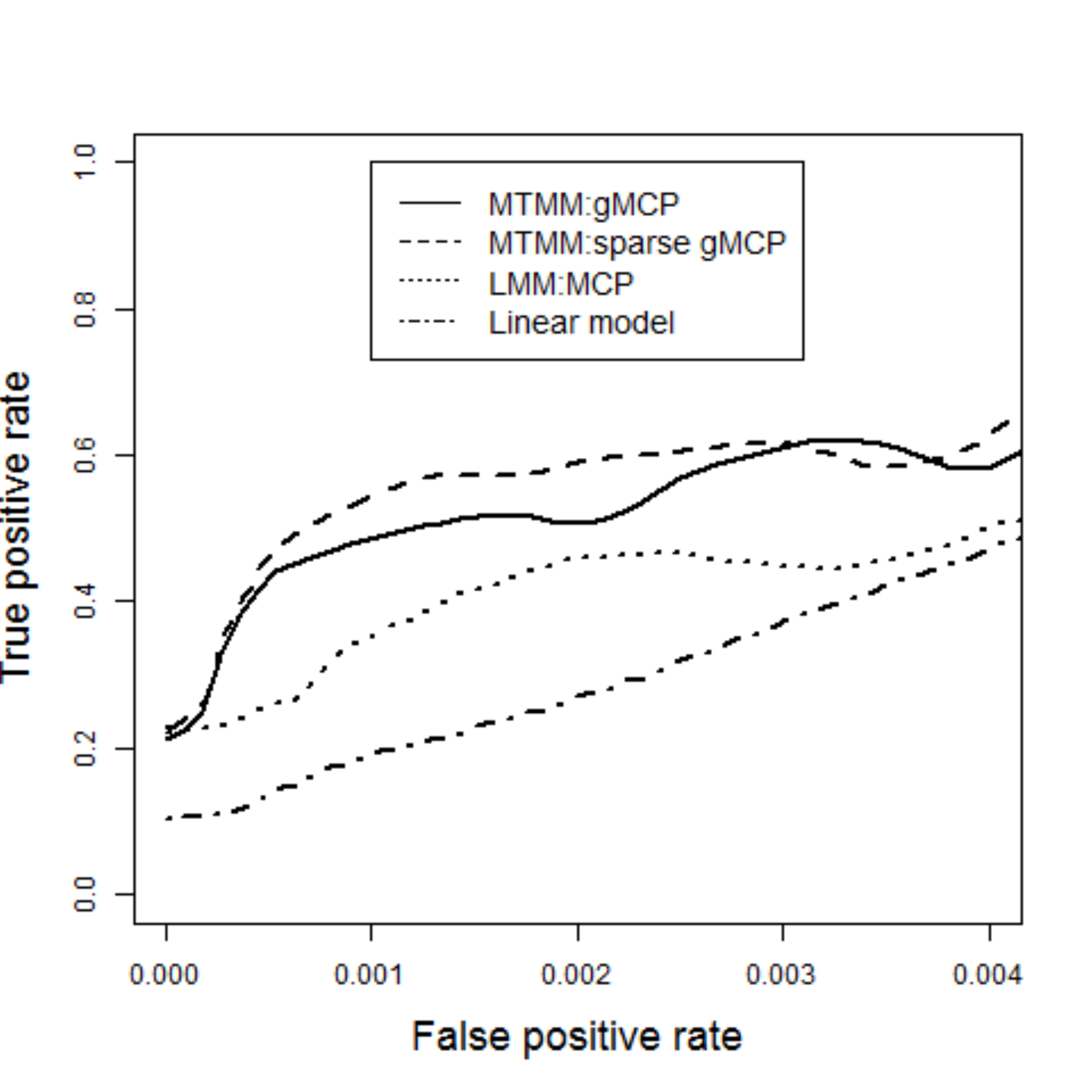}}
\centering\subfigure[Scenario 4]{\label{case4}\includegraphics[scale=0.3]{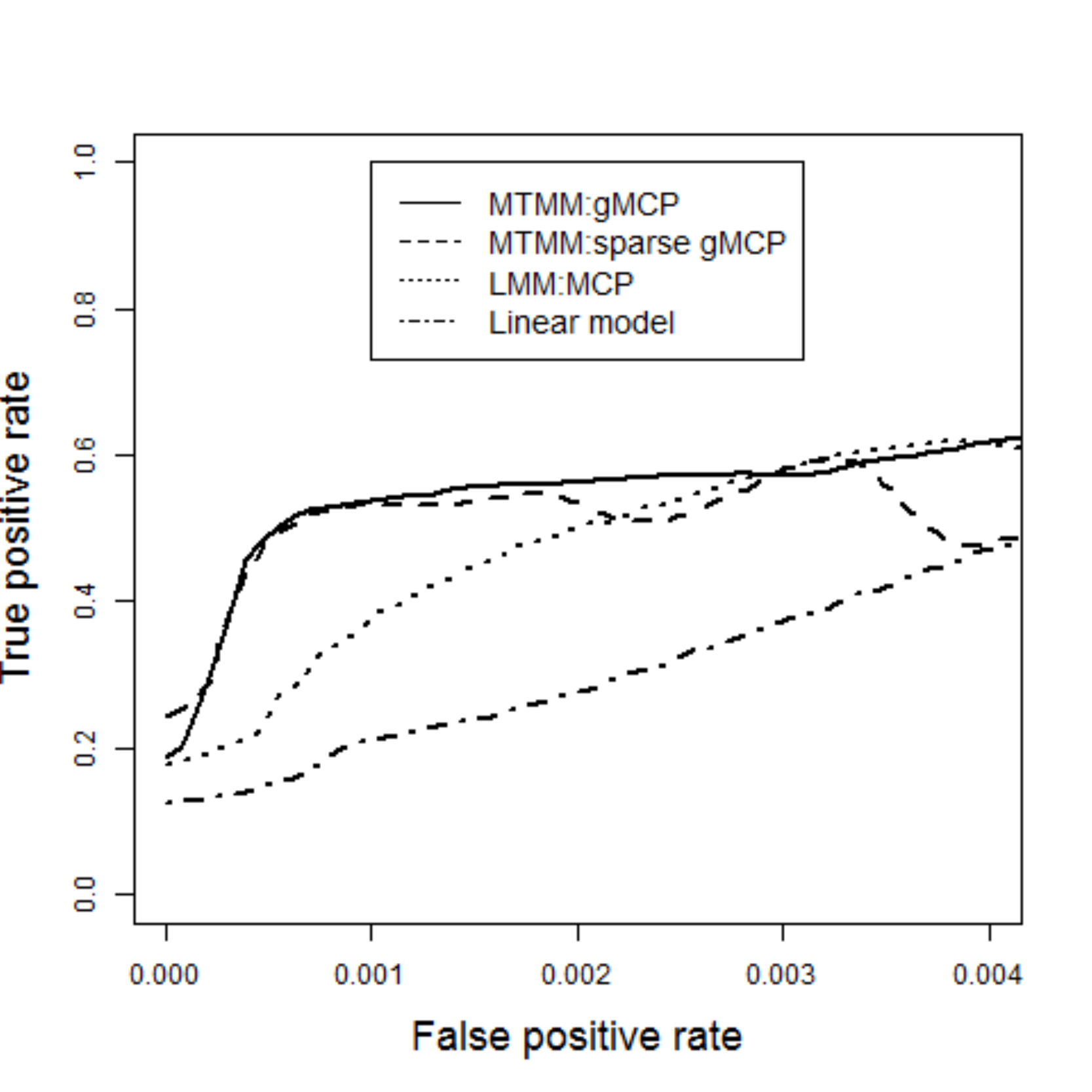}}
\centering\subfigure[Scenario 5]{\label{case5}\includegraphics[scale=0.3]{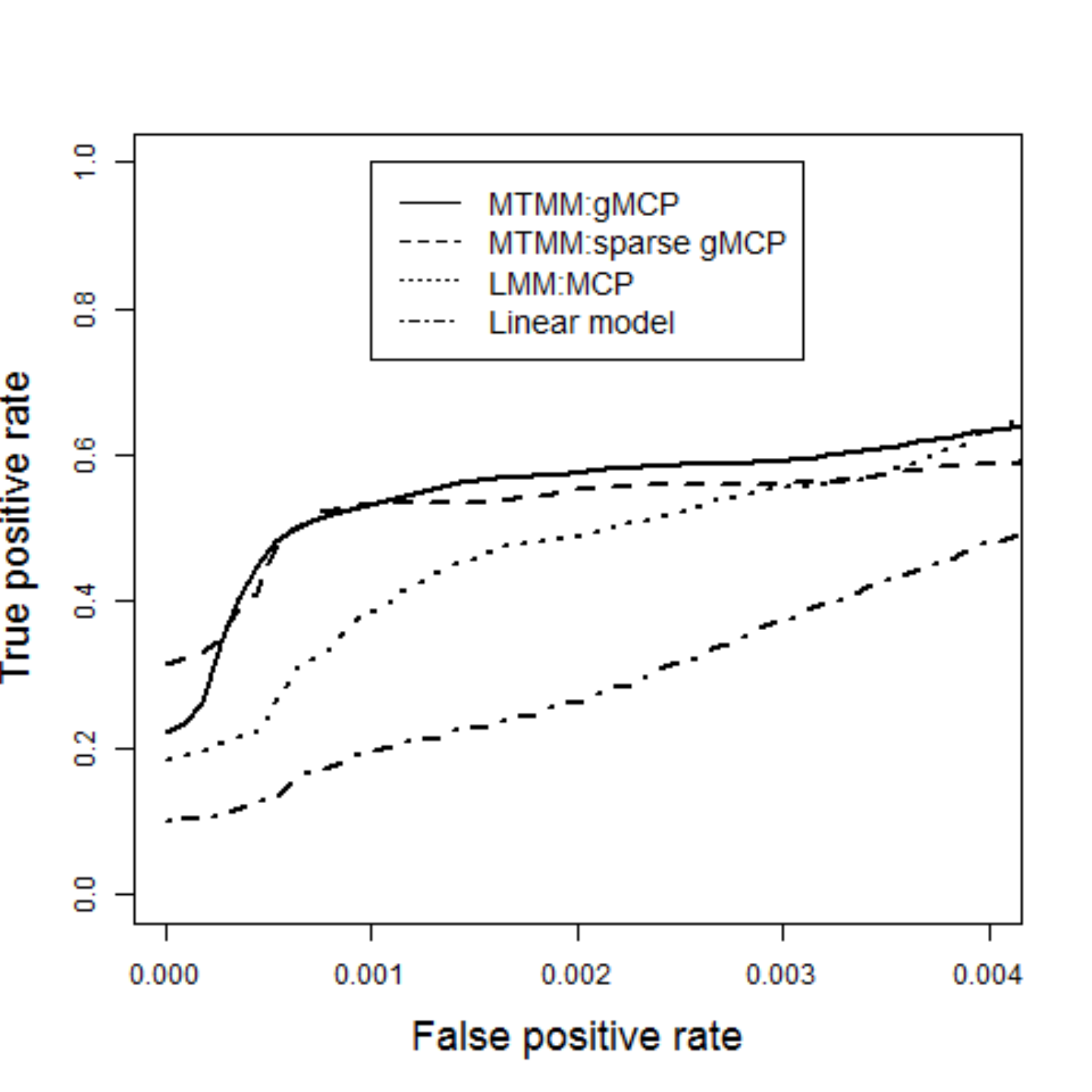}}
\centering\subfigure[Scenario 6]{\label{case6}\includegraphics[scale=0.3]{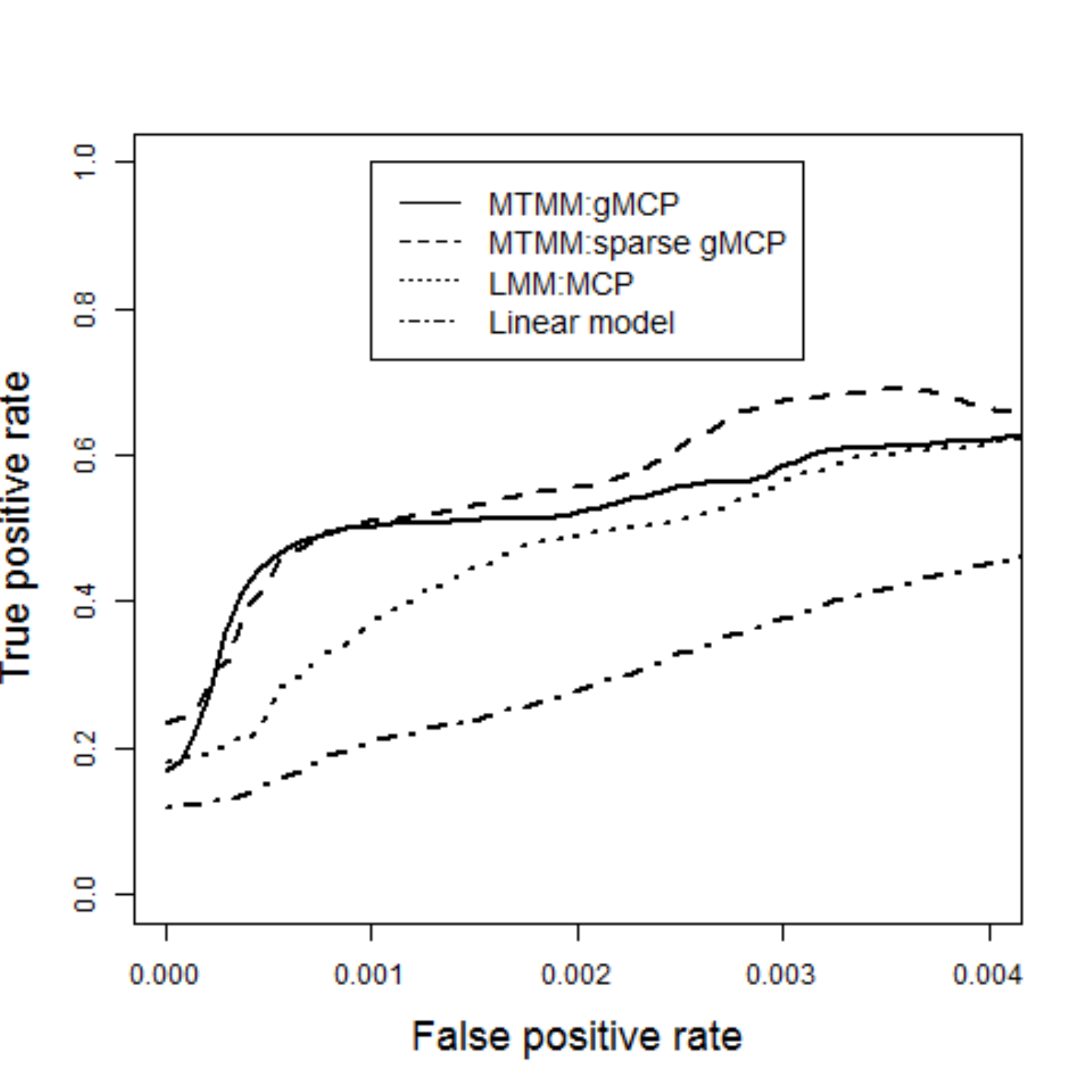}}
\caption{ROC plots for example 1--6 in homogeneity model. }\label{ROC1}
\end{figure}

\begin{figure}[ht]
\centering\subfigure[Scenario 1]{\label{case1}\includegraphics[scale=0.3]{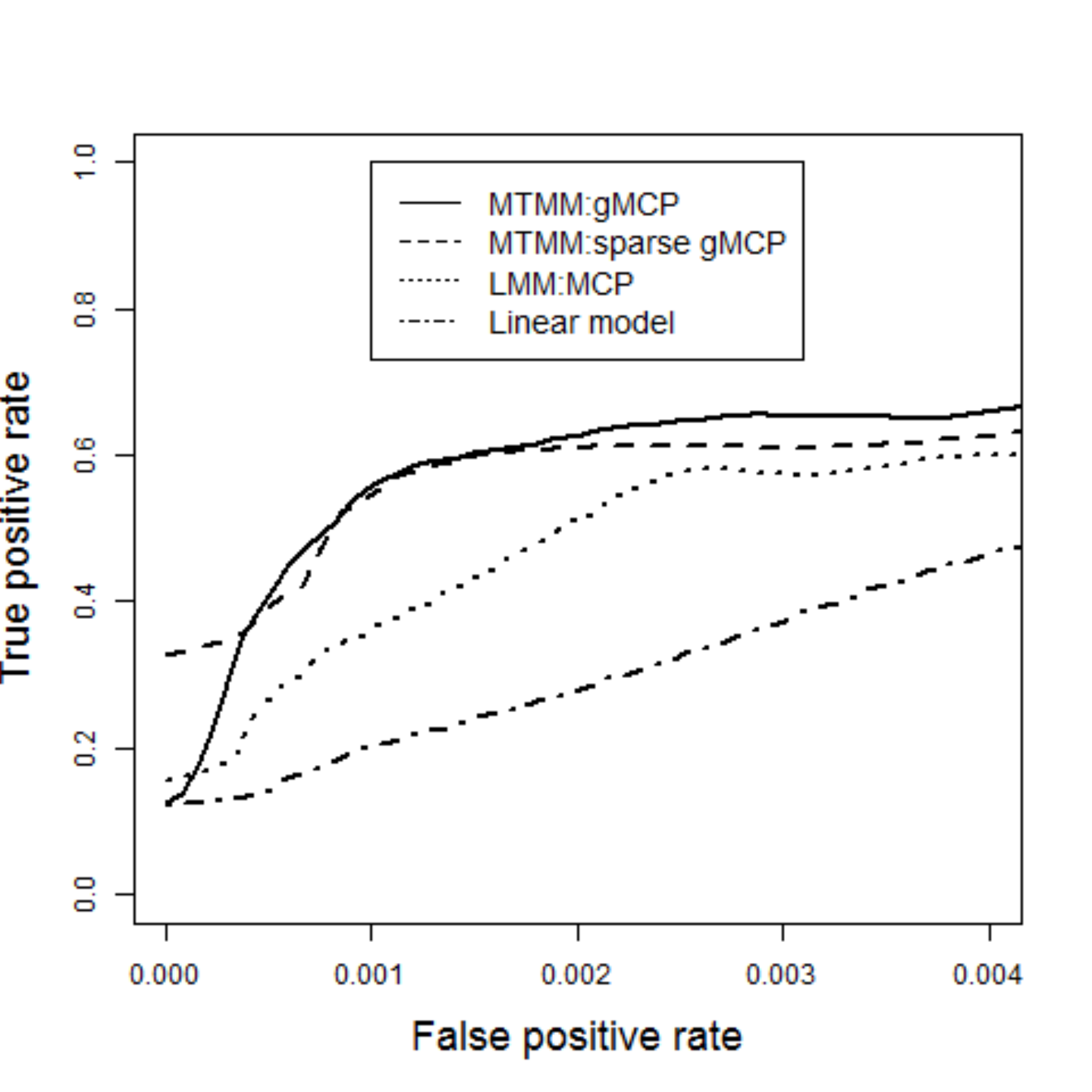}}
\centering\subfigure[Scenario 2]{\label{case2}\includegraphics[scale=0.3]{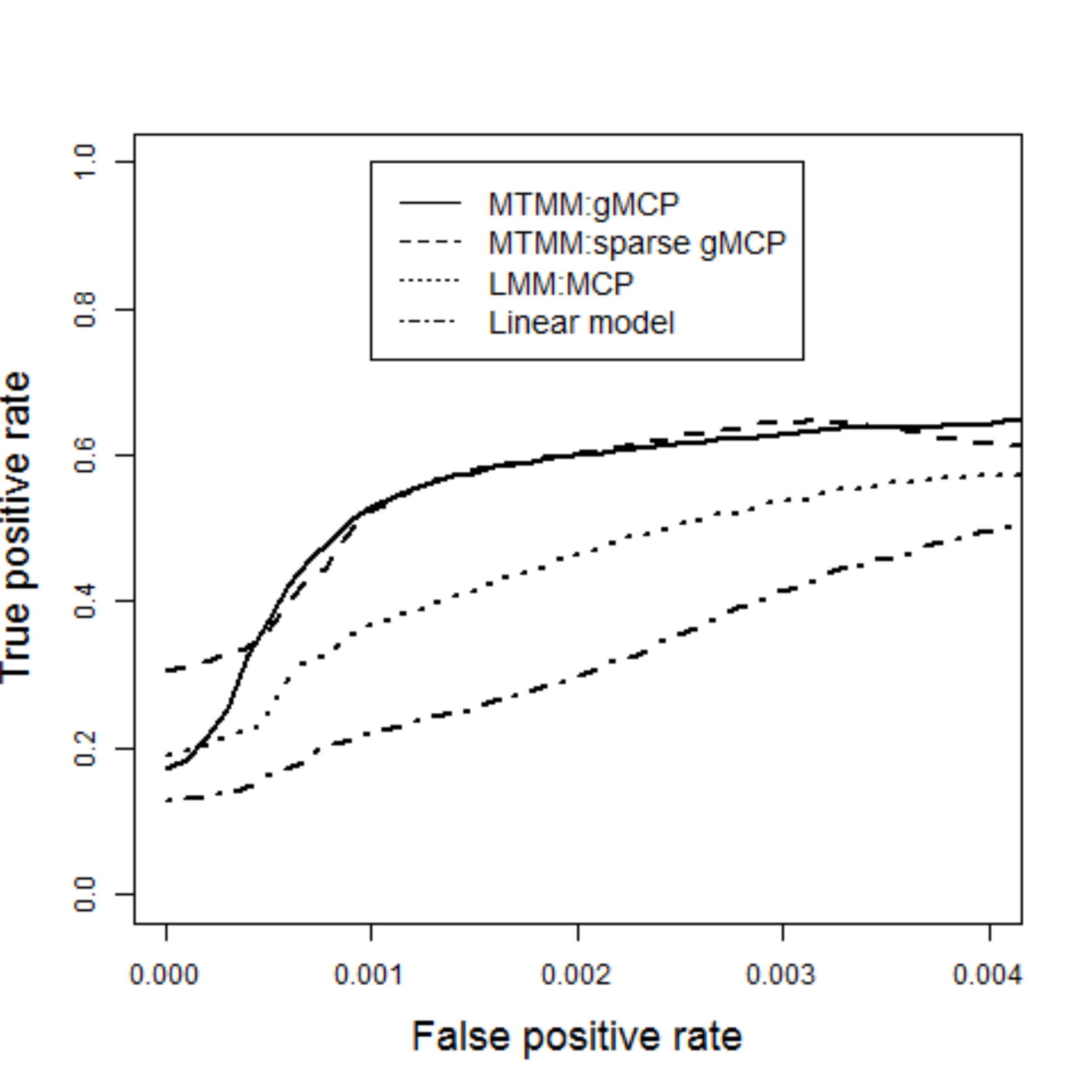}}
\centering\subfigure[Scenario 3]{\label{case3}\includegraphics[scale=0.3]{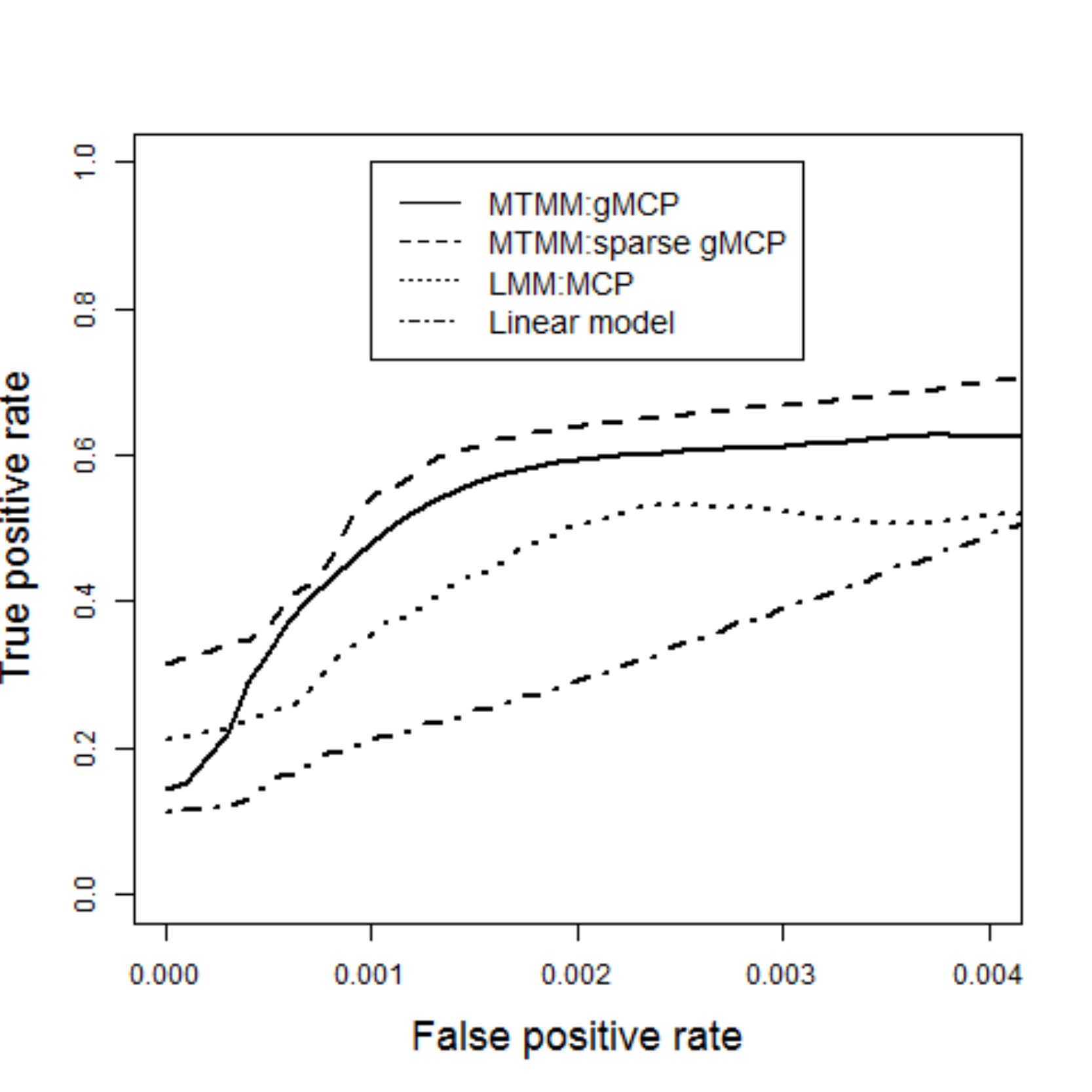}}
\centering\subfigure[Scenario 4]{\label{case4}\includegraphics[scale=0.3]{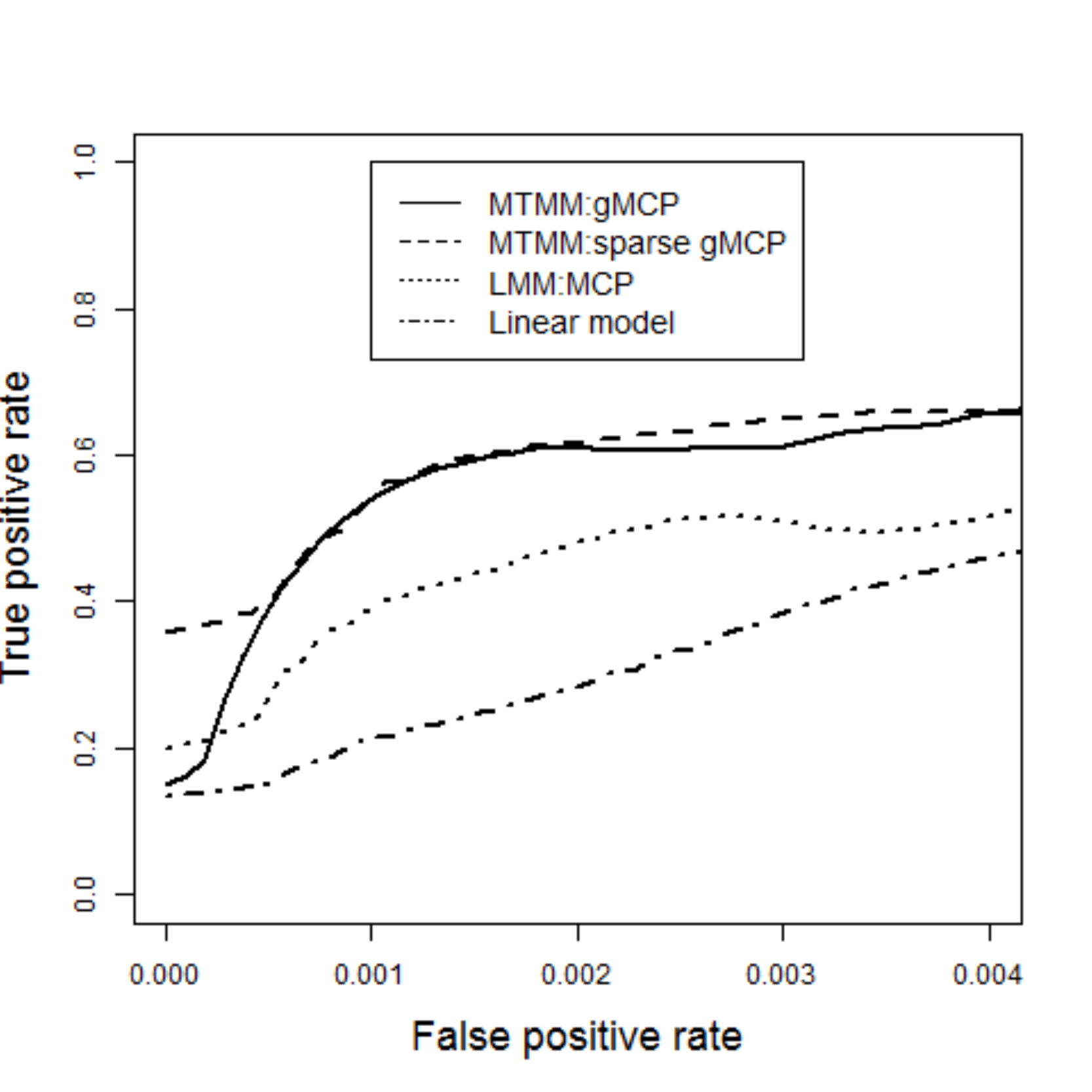}}
\centering\subfigure[Scenario 5]{\label{case5}\includegraphics[scale=0.3]{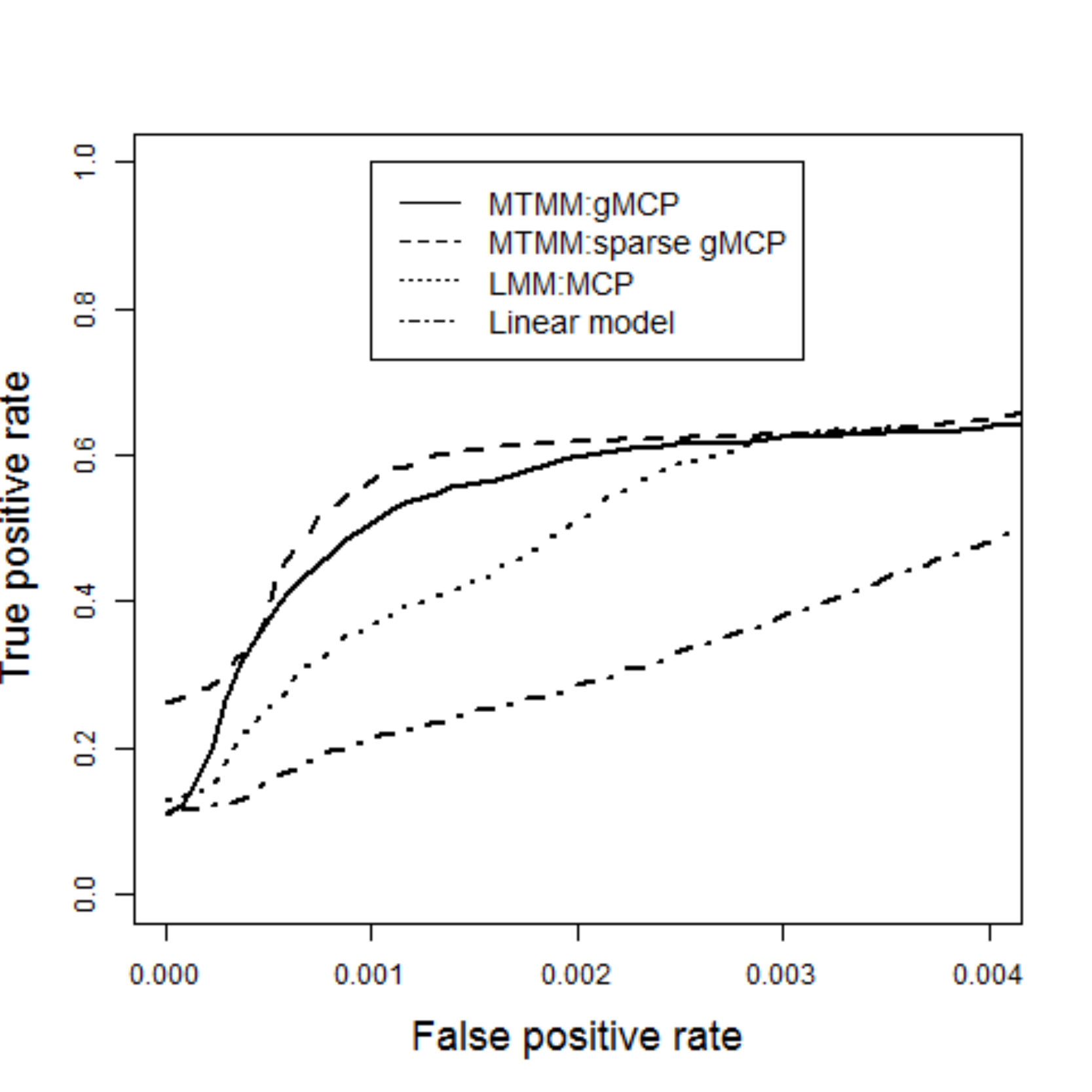}}
\centering\subfigure[Scenario 6]{\label{case6}\includegraphics[scale=0.3]{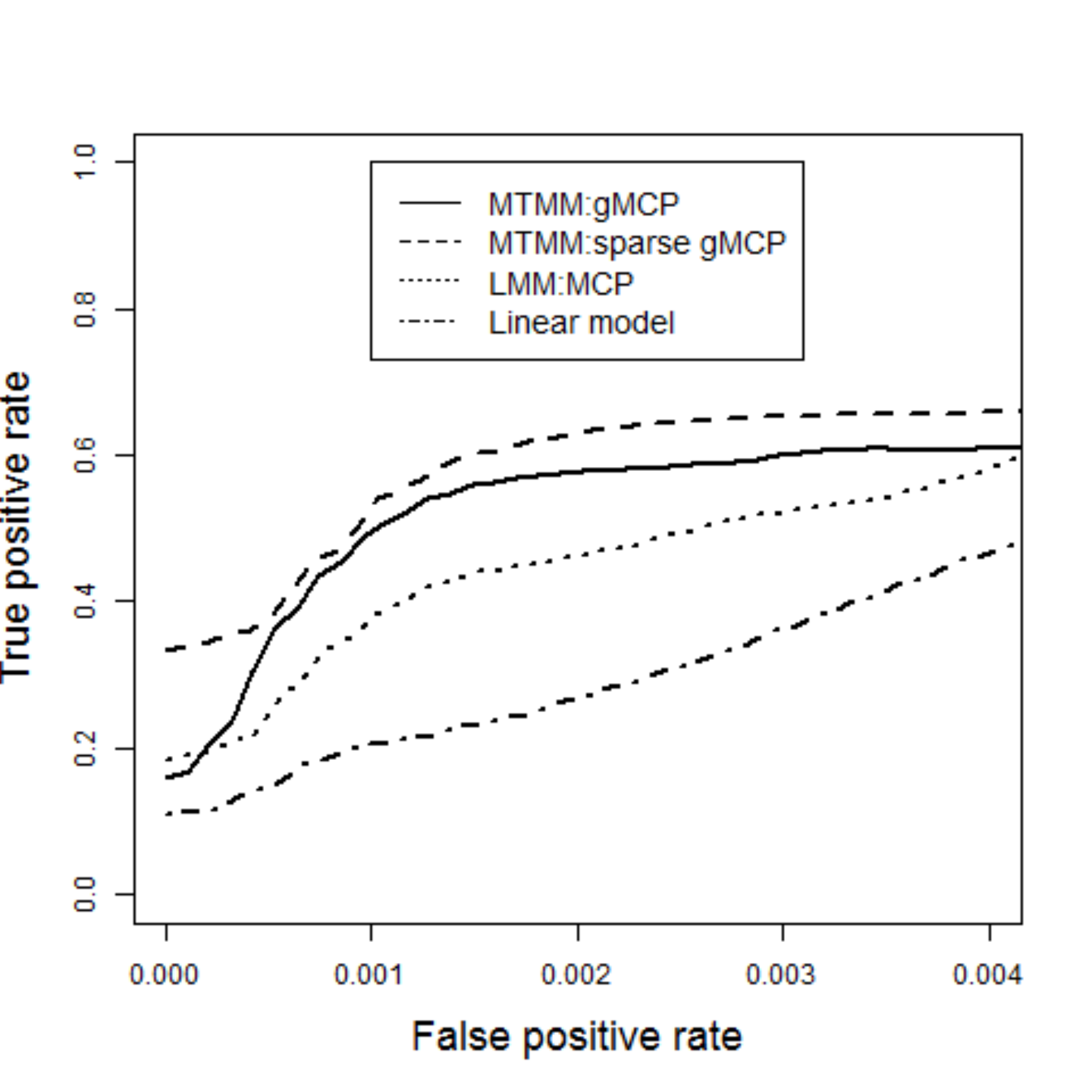}}
\caption{ROC plots for example 1--6 in heterogeneity model. }\label{ROC2}
\end{figure}

We analyze simulated data using the proposed penalized-MTMM approach on multiple traits. For comparison, we also consider penalized-LMM considering one trait at a time and linear model on each trait without consideration of variance components adjusting for relatedness among samples. For all scenarios on covariance components on both unobserved random residual and polygenic effects, ROC curves for homogeneity and heterogeneity are shown in Figure~\ref{ROC1} and~\ref{ROC2}, respectively. We also calculate the partial area under the curve (P-AUC) for each methods under each scenario~(Table~\ref{Tab:PAUC}). One can observe that under homogeneity structure, the area under the curve for group MCP and sparse group MCP using penalized-MTMM is larger than that from uni-trait penalized-LMM using MCP and uni-trait linear model. Under heterogeneity model, the P-AUC using penalized-MTMM is larger for sparse group MCP in four scenarios but also comparable in other two scenarios. The main reason for this phenomenon is that only two traits in the study. We postulate if the number of traits is going up to three or four, the improvement of sparse gMCP over gMCP in heterogeneity becomes more obvious. Furthermore, one can observe that uni-trait penalized-LMM using MCP is consistently better than univariate linear model, since uni-trait LMM-Pen takes into account the confounding relatedness in samples that is better in identifying genetic variants. To better compare prediction, we compare the multi-trait and uni-trait methods through simulation studies.

\section{Analysis of GAW 18 data}
We analyze GAW 18 data described in Section~\ref{data_description}. GAW 18 T2D--GENES study provides a GWAS data consisting of SNPs from odd autosomes. Totally, there are 472,049 SNPs over 11 autosomes for 959 samples from 20 large pedigrees. Among those SNPs, 292,355 SNPs are within the scope of gene. Using the collapsing techniques described in Section~\ref{collapse_snp}, SNPs within gene scope are collapsed into genetic scores for 10,949 genes. After quality control, 849 samples with genetic scores for 10,549 genes are used for further analysis.

The variance components for residuals are $\begin{pmatrix} 0.900 & 0.490 \\ 0.490 & 0.903\end{pmatrix}$ and the proportion between $\Sigma_d$ and $\Sigma_e$ is ($\theta$=) 0.100. One can deduce heritability for this data set that is $\theta/(1+\theta)$ (=0.091). The estimates and corresponding observed occurrence index (OOI) of the proposed approach using group MCP and sparse group MCP are shown in Table~\ref{gMCP} and~\ref{sparse_gMCP}. For comparison, we conduct uni-trait analysis using MCP, the estimates and corresponding observed occurrence index (OOI) are shown in Table~\ref{MCP}. To evaluate prediction performance, we calculate the correlation between the predictive values using BLUP in Section~\ref{prediction} and their corresponding observations. We carry out this procedure via V-fold cross-validation. The mean (sd) correlation is 0.152(0.057) and 0.186(0.084) for SBP and DBP, respectively using the proposed method on group MCP. The mean correlation is 0.139(0.070) and 0.192(0.112) for SBP and DBP, respectively using the proposed method on sparse group MCP. For comparison, The mean correlation is 0.125(0.078) and 0.146(0.074) for SBP and DBP, respectively using uni-trait method on MCP.


\begin{table}[!tpb]
\caption{Six scenarios for Covariance on both unobserved random residual and polygenic effects.}
\label{Tab:01}
\centering 
\begin{tabular}{ccc}
\hline & $\Sigma_e$ & $\Sigma_d$ \\
\hline
Scenario 1 & $\begin{pmatrix}
 0.20 & 0.04 \\
 0.04 & 0.20
\end{pmatrix}$ &
$\begin{pmatrix}
 0.40 & 0.08 \\
 0.08 & 0.40
\end{pmatrix}$\\ \hline
Scenario 2 & $\begin{pmatrix}
 0.20 & 0.10 \\
 0.10 & 0.20
\end{pmatrix}$ &
$\begin{pmatrix}
 0.40 & 0.20 \\
 0.20 & 0.40
\end{pmatrix}$\\ \hline
Scenario 3 & $\begin{pmatrix}
 0.20 & 0.16 \\
 0.16 & 0.20
\end{pmatrix}$ &
$\begin{pmatrix}
 0.40 & 0.32 \\
 0.32 & 0.40
\end{pmatrix}$\\ \hline
Scenario 4 & $\begin{pmatrix}
 0.20 & 0.04 \\
 0.04 & 0.24
\end{pmatrix}$ &
$\begin{pmatrix}
 0.40 & 0.24 \\
 0.24 & 0.40
\end{pmatrix}$\\ \hline
Scenario 5 & $\begin{pmatrix}
 0.20 & 0.16 \\
 0.16 & 0.24
\end{pmatrix}$ &
$\begin{pmatrix}
 0.40 & 0.04 \\
 0.04 & 0.40
\end{pmatrix}$\\ \hline
Scenario 6 & $\begin{pmatrix}
 0.20 & 0.16 \\
 0.16 & 0.24
\end{pmatrix}$ &
$\begin{pmatrix}
 0.40 & 0.24 \\
 0.24 & 0.40
\end{pmatrix}$\\ \hline
\end{tabular}%
\end{table}

\begin{table}[ht]
\caption{Partial AUC (standard deviation) under 6 scenarios for all methods in both homogeneity and heterogeneity models.}
\label{Tab:PAUC}
\begin{center}
\begin{tabular}{lrrrrr}
\hline
   &   Data Model & \multicolumn{2}{c}{MTMM} & LMM & Linear Model\\
Penalty  &  & gMCP & sparse gMCP & MCP & MCP\\
\hline
Scenario 1 & Homogeneity & 0.537(0.134) & 0.484(0.140) & 0.355(0.084) & 0.344(0.062) \\
Scenario 2 & Homogeneity & 0.486(0.128) & 0.468(0.115) & 0.340(0.087) & 0.372(0.067) \\
Scenario 3 & Homogeneity & 0.411(0.131) & 0.452(0.125) & 0.359(0.092) & 0.355(0.071) \\
Scenario 4 & Homogeneity & 0.517(0.105) & 0.471(0.130) & 0.332(0.108) & 0.359(0.063) \\
Scenario 5 & Homogeneity & 0.520(0.089) & 0.487(0.102) & 0.337(0.094) & 0.367(0.063) \\
Scenario 6 & Homogeneity & 0.449(0.128) & 0.460(0.128) & 0.324(0.100) & 0.357(0.055) \\
Scenario 1 & Heterogeneity & 0.540(0.089) & 0.499(0.095) & 0.351(0.082) & 0.360(0.065) \\
Scenario 2 & Heterogeneity & 0.531(0.101) & 0.501(0.106) & 0.352(0.083) & 0.375(0.062) \\
Scenario 3 & Heterogeneity & 0.558(0.093) & 0.574(0.108) & 0.324(0.102) & 0.366(0.063) \\
Scenario 4 & Heterogeneity & 0.510(0.092) & 0.521(0.104) & 0.333(0.091) & 0.350(0.072)\\
Scenario 5 & Heterogeneity & 0.489(0.103) & 0.515(0.122) & 0.324(0.095) & 0.357(0.070) \\
Scenario 6 & Heterogeneity & 0.518(0.093) & 0.558(0.090) & 0.327(0.084) & 0.342(0.058) \\

\hline
\end{tabular}
\end{center}
\end{table}


\begin{table}[ht]
\caption{Gene selected by penalized-MTMM using gMCP.}
\label{gMCP}
\begin{center}
\begin{tabular}{lrrrlrrr}
  \hline
Gene & Trait1 & Trait2 & OOI & Gene & Trait1 & Trait2 & OOI \\
  \hline
DFFA & 0.001 & 0.000 & 0.400 & NCAM1 & -0.017 & -0.005 & 0.690 \\
  LOC390998 & 0.031 & 0.024 & 0.900 & OR8A1 & -0.002 & -0.030 & 0.970 \\
  MOBKL2C & -0.009 & -0.002 & 0.440 & FNDC3A & 0.008 & 0.015 & 0.650 \\
  SLC16A4 & -0.007 & -0.000 & 0.550 & INOC1 & 0.029 & 0.026 & 0.920 \\
  LCE1A & 0.018 & 0.016 & 0.750 & PIAS1 & 0.002 & 0.001 & 0.340 \\
  PYCR2 & 0.004 & 0.001 & 0.500 & TLCD2 & 0.004 & -0.001 & 0.520 \\
  ALCAM & -0.002 & -0.002 & 0.440 & HS3ST3B1 & 0.012 & 0.004 & 0.660 \\
  EIF2B5 & -0.001 & -0.002 & 0.460 & FLII & -0.004 & -0.005 & 0.360 \\
  ZCCHC10 & 0.019 & 0.011 & 0.680 & RAMP2 & -0.004 & -0.011 & 0.540 \\
  ANKHD1 & 0.007 & 0.004 & 0.520 & ANKRD40 & -0.002 & 0.022 & 0.850 \\
  LOC100130230 & -0.003 & -0.005 & 0.540 & C19orf38 & -0.009 & 0.025 & 0.970 \\
  PAPOLB & 0.015 & 0.016 & 0.820 & NDUFA13 & -0.002 & 0.001 & 0.370 \\
  RPA3 & -0.032 & -0.010 & 0.890 & ZNF826 & -0.006 & 0.020 & 0.900 \\
  ELMO1 & -0.008 & -0.001 & 0.430 & SYT3 & 0.016 & 0.011 & 0.800 \\
  OGDH & 0.040 & 0.037 & 0.990 & ZNF611 & -0.008 & -0.017 & 0.770 \\
  EXOSC2 & 0.029 & 0.040 & 0.990 & AIRE & -0.004 & -0.011 & 0.570 \\
  LOC390084 & -0.008 & -0.007 & 0.510 &  \\
   \hline
\end{tabular}
\end{center}
\end{table}

\begin{table}[ht]
\caption{Gene selected by penalized-MTMM using sparse gMCP.}
\label{sparse_gMCP}
\begin{center}
\begin{tabular}{crrrrcrrrr}
\hline
Gene & \multicolumn{2}{c}{Trait 1} & \multicolumn{2}{c}{Trait 2} & Gene & \multicolumn{2}{c}{Trait 1} & \multicolumn{2}{c}{Trait 2} \\
  \cline{2-5}\cline{7-10}
     & Est. & OOI &  Est.   & OOI &   & Est. & OOI & Est. & OOI \\
  \hline
AIRE & -0.004 & 0.020 & -0.007 & 0.450 & MOBKL2C & -0.006 & 0.510 &  &  \\
  ANKRD40 & 0.001 & 0.580 & 0.005 & 0.580 & NCAM1 & -0.012 & 0.750 &  &  \\
  C19orf38 & -0.004 & 0.690 & 0.010 & 0.690 & OGDH & 0.024 & 0.810 & 0.031 & 0.980 \\
  CSF1 & -0.005 & 0.030 & -0.008 & 0.490 & OR8A1 & -0.004 & 0.770 & -0.016 & 0.770 \\
  EIF2B5 & -0.007 & 0.000 & -0.010 & 0.490 & PAPOLB & 0.010 & 0.200 & 0.015 & 0.610 \\
  ELMO1 & -0.009 & 0.610 &  &  & PYCR2 & 0.004 & 0.510 &  &  \\
  EXOSC2 & 0.027 & 0.000 & 0.042 & 1.000 & RAMP2 & -0.005 & 0.050 & -0.009 & 0.570 \\
  FLII & -0.004 & 0.000 & -0.006 & 0.550 & RPA3 & -0.018 & 0.810 & -0.001 & 0.350 \\
  FNDC3A & 0.017 & 0.000 & 0.026 & 0.710 & SCYL1BP1 & -0.001 & 0.530 &  &  \\
  HS3ST3B1 & 0.004 & 0.540 & 0.000 & 0.290 & SLC16A4 & -0.016 & 0.760 &  &  \\
  INOC1 & 0.009 & 0.560 & 0.010 & 0.600 & TLCD2 & 0.007 & 0.650 &  &  \\
  LCE1A & 0.001 & 0.320 & 0.001 & 0.420 & ZNF611 & -0.014 & 0.020 & -0.022 & 0.880 \\
  LOC100130230 & -0.007 & 0.000 & -0.011 & 0.700 & ZNF826 & -0.001 & 0.540 & 0.003 & 0.540 \\
  LOC390998 & 0.013 & 0.700 & 0.012 & 0.700 & \\
   \hline
\end{tabular}
\end{center}
\end{table}

\begin{table}[ht]
\caption{Gene selected by uni-trait LMM using MCP separately on each trait.}
\label{MCP}
\begin{center}
\begin{tabular}{crrrrcrrrr}
\hline
Gene & \multicolumn{2}{c}{Trait 1} & \multicolumn{2}{c}{Trait 2} & Gene & \multicolumn{2}{c}{Trait 1} & \multicolumn{2}{c}{Trait 2} \\
  \cline{2-5}\cline{7-10}
     & Est. & OOI &  Est.   & OOI &   & Est. & OOI & Est. & OOI \\
  \hline
ANKHD1 & 0.017 & 0.710 &  &  & NCAM1 & -0.024 & 0.800 &  &  \\
  ARSB & -0.007 & 0.490 &  &  & OGDH & 0.022 & 0.730 & 0.011 & 0.760 \\
  C1orf128 & 0.015 & 0.680 &  &  & PIAS1 & 0.016 & 0.700 &  &  \\
  C3orf20 & 0.007 & 0.500 &  &  & PYCR2 & 0.008 & 0.660 &  &  \\
  DFFA & 0.005 & 0.450 &  &  & RPA3 & -0.030 & 0.900 &  &  \\
  HS3ST3B1 & 0.022 & 0.810 &  &  & SFRS12 & 0.007 & 0.580 &  &  \\
  INOC1 & 0.019 & 0.760 & 0.004 & 0.540 & SYT3 & 0.024 & 0.750 &  &  \\
  LCE1A & 0.007 & 0.430 &  &  & ZCCHC10 & 0.027 & 0.780 &  &  \\
  LOC390998 & 0.028 & 0.830 &  &  & EXOSC2 &  &  & 0.038 & 0.980 \\
  MCM7 & -0.001 & 0.410 &  &  & FNDC3A &  &  & 0.015 & 0.720 \\
  MKNK1 & -0.002 & 0.270 &  &  & ZNF611 &  &  & -0.011 & 0.750 \\
   \hline
\end{tabular}
\end{center}
\end{table}

\section{Discussion}
We have presented a penalized multi-trait mixed model (Penalized-MTMM) for detecting pleiotropic genetic associations among multiple traits in the presence of pedigree structure. The approach combines the advantages of mixed models that allow for elegant correction for pedigree-based family data, integrative analysis of multiple traits that borrow strengths across traits and joint multi-variant models that take the joint effects of sets of genetic variants into account rather than one single variant at a time. In the joint multi-variant models, we consider both homogeneity and heterogeneity structure using group MCP and sparse group MCP, respectively. We use ROC to evaluate selection performance for penalized-MTMM comparing with penalized-LMM considering one trait at a time and a linear model. To evaluate prediction performance, we use BLUP to find the predictive values and the correlations of the predictive values and their corresponding observations are calculated subsequently. Our numerical studies show that the proposed approach has satisfactory performance.

Confounder effects and population structure induce spurious correlations between genotype and phenotype, complicating the genetic analysis. Mixed models accounting for the presence of such structure are well studied and have been shown to greatly reduce the impact of this confounding source of variability. For instance, EIGENSTRAT was built upon the idea of extracting the major axes of population differentiation using a PCA decomposition of the genotype data and subsequently including them into the model as additional covariates~\citep{price2006principal}. The penalized-MTMM can consider both of confounder effects and population structure depending the choice of random effects.
On the other hand, mixed models also show its strength in coping with repeated measures in longitudinal studies. The penalized-MTMM can handle data from longitudinal studies with multiple traits.

In the similar fashion, our method can be applied to conduct integratively analysis of multiple GWAS with correlated traits.
\bibliographystyle{plainnat}
\bibliography{reference}

%

\end{document}